\documentclass[aps,pra,superscriptaddress,amsmath,amssymb,preprintnumbers,floatfix,showpacs,11pt]{revtex4}

\usepackage{amssymb}
\usepackage{epsfig}

\begin{document}
\title{ Quantum properties of the parametric amplifier
with and without pumping field fluctuations }
\author{ Faisal A. A. El-Orany }

\author{  J.Pe\v{r}ina
 }

 \affiliation{ Department of Optics, Palack\'y University, 17.~listopadu~50,
772 07~Olomouc, Czech Republic}

\author{M. Sebawe Abdalla}

\affiliation{Mathematics Department, College of Science, King Saud
University, P.O. Box 2455, Riyadh 11451, Saudi Arabia}

\date{\today}

\begin{abstract}
The parametric amplifier with and without the pumping fluctuations
of coupling function is considered when the  fields are initially
prepared  in coherent light.
 The pumping fluctuations are assumed to be normally
distributed with time-dependent variance. The effects of
antibunching and anticorrelation of photons on the photon
distribution, correlation between modes and  factorial moments
  are demonstrated.
A possible enhancement of photon antibunching for certain values
of initial mean photon numbers is  shown and discussed. We have
shown also that new states (called modified squeezed vacuum states
or even thermal states) can be generated from such  an
interaction. Further, we have demonstrated that the sum
photon-number distribution can exhibit collapses and revivals in
the photon-number domain  somewhat similar to those known in the
Jaynes-Cummings model.

\end{abstract}

\pacs{42.50Dv,42.60.Gd}
   \maketitle

\section{Introduction}

Quadratic parametric processes have been considered since 1961,
when a quantum-mechanical model of the parametric amplification
and frequency conversion, treating the pump field classically and
neglecting the losses, had been proposed in \cite{in1}.
Among these processes parametric amplifier takes a considerable
interest since degenerate and nondegenerate parametric amplifiers can
 perfectly provide single-mode and two-mode squeezing
(squeezed light has less noise than a coherent light
in one of the field quadratures and can exhibit a number of features
having no classical analogue).
Further,  parametric amplifier  has been employed in experiments. For
example, the
fourth-order interference effects arise when pairs of photons produced in
 parametric amplifier are injected into Michelson interferometers \cite{kwia}.
The second-order interference is observed in the superposition of
signal photons from two coherently pumped parametric amplifiers when the
paths of the idler photons are aligned \cite{zou1}.

It is well-known that the interaction of light with matter
changes the statistical properties of light. The nature of the
change depends on the kind of interaction process and on the
statistical properties of the incident electromagnetic field.
An investigation of the statistical properties
 of the nondegenerate parametric amplifier
 (without losses) has been performed using different
techniques, such as Heisenberg technique \cite{in2} and Schr\"{o}dinger
technique \cite{in3}. The main attention in these articles has been
paid to the single mode case, i.e. only one mode (signal or idler) is
detected. The anticorrelation in this model is an interesting effect
\cite{in4}, where  the variance of the photon number is less than the average
photon number and the photocounting distribution becomes narrower than the
corresponding Poisson distribution for a coherent state with the same mean
photon number. These nonclassical phenomena are established for certain
values of the  phase of incident fields.
On the other hand, the quantum
statistical properties of the parametric amplification and
generation including losses in the medium have been studied by
means of the generalized Fokker-Planck equation for the
distribution related to the antinormal ordering \cite{in5}
 showing that the normal counting generating function
decomposes into the product of two functions which are the
generating functions for the superposition of coherent and
chaotic fields and that the outgoing field exhibits the
anticorrelation effect.
Further, it is convenient mentioning that the photon counting distribution
together with the waiting-time distribution  techniques have
been used to analyze the
nonclassical effects for light beams generated by the lossy  degenerate
optical parametric amplifier \cite{add}.

Finally, we refer to parametric processes  with stochastic pumping
\cite{in6,{in7},{in8},{in9},{in10},{in11}} modelled by means of a
stochastic differential  equations. One can find the total
photon number, as a function of time, of frequency up-conversion
process has a turning point and the antibunching effect is
predicted regardless of the mean photon numbers available in the
process. Also the effect of pumping fluctuations are considered in
\cite{in9,{in10},{in11}}, where the observed
sub-Poissonian behaviour is progressively degraded as the level
of fluctuations in the pumping radiation increases.
This means that the pump fluctuations play a similar role as reservoir, where
the interaction with the environment leads to the
destruction of the nonclassical effect, i.e. to the coherent energy extraction
from the quantum mechanical system.

In the present paper we study the quantum properties of the
 parametric amplifier with the classical pumping by considering
the pump fluctuations in the system. More illustratively,
the lossless effective Hamiltonian \cite{in9,{in10},{in11}} controlling
 the system is

\begin{equation}
\hat{H}=\hbar \omega_{1}\hat{a}^{\dagger}_{1}\hat{a}_{1}
+\hbar \omega_{2}\hat{a}^{\dagger}_{2}\hat{a}_{2}
-\hbar g[\hat{a}_{1}\hat{a}_{2}\exp(i\omega t-i\phi) +{\rm h.c}]
,\label{1}
\end{equation}
where $\hat{a}_{j},j=1,2$, are the annihilation operators
designated  to the signal and idler mode, respectively;
$\omega_{j}$ are the natural  frequencies of oscillations
of the uncoupled modes with $\omega =\omega_{1}+\omega_{2}$ (assuming
the resonance frequency condition holds); $\phi$ is the initial phase of
the pump and {\rm h.c.} is the Hermitian
conjugate; $g$ is the real coupling constant (phase matching is assumed)
and we assume $g=g_{0}+\varepsilon g_{1}$, that
is the coupling constant consists of two parts: $g_{0}$ is the deterministic
part and $g_{1}$ is the correction part
(i.e. $| g_{0}| >> |\varepsilon  g_{1}|$), and $\varepsilon$ is a real
fluctuating quantity obeying time-dependent  Gaussian distribution
described by

\begin{equation}
P(\varepsilon,t)=\frac{1}{\sqrt{2\pi\sigma(t)}}
\exp \left[-\frac{\varepsilon^{2}}{2\sigma(t)}\right]
,\label{1a}
\end{equation}
where the  standard  deviation is $\sigma(t) =\sigma_{0}[1-\exp(-\mu t)]$,
$\sigma_{0}$ is the maximum standard  deviation and $\mu$ is a finite real
parameter.
This is the difference from what has been done earlier
\cite{in9,{in10},{in11}}, where the standard deviation is time independent.
In this way pump fluctuations are taken into account.
Indeed, time-dependent standard  deviation can lead to the
enhancement of nonclassical effects
compared to the time-independent case, as we shall see later.
For any observable  $\hat{F}(\varepsilon,t)$,
the average processs will be performed as follows:

\begin{equation}
\langle \langle \hat{F}(\varepsilon,t)\rangle \rangle=
\int_{-\infty}^{\infty} \langle\hat{F}(\varepsilon,t)\rangle_{\rm f}
P(\varepsilon,t)d\varepsilon
,\label{1b}
\end{equation}
where the two angle brackets means that there are two average operations, one
is the average (expectation value) over  the initial states of the
field $\langle ..\rangle_{\rm f}$ and the other is the average over the stochastic parameter $\varepsilon$
using  the distribution (\ref{1a}) in the sense described  on the right-hand
side of (\ref{1b}). Such an averaging process will be frequently used in this
article.

We should stress here that the model will be treated generally for two cases
with and without the pump fluctuations where the comparison is instructive.
Also we give new insight into the possibility to generate nonclassical light
based on the change of intensity of initial incident fields which has not been
considered earlier \cite{in5,{in9},{in10},{in11}}.

For completeness, the solution of the Heisenberg's equations
for the Hamiltonian (\ref{1}) is well known

\begin{eqnarray}
\begin{array}{lr}
\hat{a}_{1}(t)=\exp(-i\omega_{1} t)[\hat{a}_{1}(0) \cosh (gt)
+i\hat{a}^{\dagger}_{2}(0) \sinh (gt) \exp (i\phi)], \\
\\
\hat{a}_{2}(t)=\exp(-i\omega_{2} t)[\hat{a}_{2}(0) \cosh (gt)
+i\hat{a}^{\dagger}_{1}(0) \sinh (gt) \exp (i\phi)].
\label{2}
\end{array}
\end{eqnarray}

The small term $ |\varepsilon|  g_{1}<<| g_{0}|$ is considered as a small
perturbation  and the standard deviation
$\sigma(t)$ is assumed as a slowly varying quantity compared to the
variations in (\ref{2}).

This article is organized as follows: In section 2 the basic
equations and
relations such as anticorrelation functions,  reduced factorial moments
and sum photon-number distribution are given when the modes are
initially prepared in coherent light.
In section 3 discussion of the results is
performed and the conclusions are summed in section 4.

\section{Basic relations and equations}

As it is well known light fields for which the $P$ representation is  a
well-behaved distribution cannot exhibit nonclassical features \cite{in12,{in13}}.
$P$-function
for the single mode   as output from  parametric amplifier (\ref{1})
is well defined, because in the parametric systems
the evolution broadens the single mode $P$ distribution
due to the spontaneous pump photon decay \cite{band}.
This also increases the radius of the Wigner contour
compared to the initial one \cite{ba1} and reflects that there is no
(single mode) nonclassical behaviour.
For this reason we focus our attention on the behaviour of the
compound modes when the modes  are initially prepared in coherent states.

Now, cross-correlation may be used to
describe anticorrelation between modes, which may cause that the
variance of the photon number is less than the average of the photon
number giving then antibunching; it can be measured by detecting
single modes separately by two photodetectors and correlating their outputs.
Cross-correlation between  mode 1 and mode 2 is controlled by

\begin{equation}
\langle :\triangle W_{1}\triangle W_{2}:\rangle =\langle \langle:\hat{a}_{1}^{\dagger
}(t)\hat{a}
_{1}(t)\hat{a}_{2}^{\dagger }(t)\hat{a}_{2}(t):\rangle \rangle-
\langle \langle\hat{a}
_{1}^{\dagger }(t)\hat{a}_{1}(t)\rangle \langle \hat{a}_{2}^{\dagger
}(t)\hat{a}_{2}(t)\rangle\rangle , \label{2a}
\end{equation}
where $W_{j}=\hat{a}_{j}^{\dagger }(t)\hat{a}_{j}(t), j=1,2$,
and : : denotes the normally ordered operator, i.e. creation operators
$\hat{a}^{\dagger}_{j} $ are to the left of annihilation operators
$\hat{a}_{j}$;  the double angle brackets in
the right-hand side has the same meaning as before.

For input coherent light ($|\alpha_{1},\alpha_{2}\rangle$) equation (\ref{2a})
together with (\ref{2}) reduce to \cite{in5,{in9},{in10},{in11}}
\begin{eqnarray}
\langle :\triangle W_{1}\triangle W_{2}:\rangle =
\Big\langle \frac{1}{4}(2|\alpha_{1}|^{2}+
2|\alpha_{2}|^{2}+1)\sinh^{2}(2gt)
\nonumber\\
-\frac{1}{2}|\alpha_{1}||\alpha_{2}| \sinh (4gt) \sin \psi
\Big\rangle ,\label{2cc}\\
=
 \frac{1}{8}(2|\alpha_{1}|^{2}+ 2|\alpha_{2}|^{2}+1)
 \left\{ \cosh(4g_{0}t)\exp[8\sigma(t)g^{2}_{1}t^{2}]-1\right\}
\nonumber \\
-\frac{1}{2}|\alpha_{1}||\alpha_{2}| \sinh (4g_{0}t)
\exp[8\sigma(t)g^{2}_{1}t^{2}] \sin \psi
 , \label{2c}
\end{eqnarray}
where
$\alpha_{j}=|\alpha_{j}|
\exp(i\phi_{j}),j=1,2$ ( $|\alpha_{j}|$ and $\phi_{j}$ are the amplitude
and the phase of the initial coherent $j$th mode (signal or idler)),
and $\psi=\phi_{1}+\phi_{2}-\phi$ represents the difference between the
phase-sum of the initial signal and idler modes and the initial phase of
the pump, which is the quantity determining the phase mismatch.
In (\ref{2cc}) and in the following we have performed the average with
respect to the initial coherent states and the average  over stochastic
 pumping field,  which is indicated
by the angle brackets (in
(\ref{2c})  the integration over $\varepsilon$ has been carried out using
the distribution (\ref{1a}) after prescription (\ref{1b})).

We proceed by writing down  the formulae of the reduced factorial moments
and sum photon-number
distribution for compound modes; the details of the derivation
for these relations can be found in  \cite{in9,{in10}}, respectively as
\vspace{-.2cm}
\begin{eqnarray}
\langle W^{k}\rangle
=\Big\langle \sum_{l=0}^{k}
{k \choose l}
\lambda^{k-l}_{1}
\lambda^{l}_{2}
{\rm L}_{k-l}(-\frac{A_{1}}{\lambda_{1}})
{\rm L}_{l}(-\frac{A_{2}}{\lambda_{2}}) \Big\rangle ,\label{11}\\
=\sum_{l=0}^{k}\sum_{l_{1}=0}^{k-l}\sum_{l_{2}=0}^{l}\sum_{l_{3}=0}^{k-l_{1}-l_{2}}
{k \choose l}{k-l \choose l_{1}}{l \choose l_{2}}\nonumber\\
{k-l_{1}-l_{2} \choose l_{3}}\frac{(-1)^{k+l_{3}-l-l_{1}} (k-l)! l!
h^{l_{1}}_{1} h^{l_{2}}_{2}}
{l_{1}!l_{2}!} \nonumber\\
\times (\frac{1}{2})^{k-l_{1}-l_{2}} \exp [2g_{0}t(l-l_{1}-l_{3})
+2\sigma(t) g^{2}_{1}t^{2}
(l-l_{1}-l_{3})^{2}], \label{11a}
\end{eqnarray}
\vspace{-1.5cm}
\begin{eqnarray}
P(n,t)
=\Big\langle
\frac{
\exp\left[ -\frac{A_{1}}{(1+\lambda_{1})}
-\frac{A_{2}}{(1+\lambda_{2})}\right]
}{(1+\lambda_{1})(1+\lambda_{2})}
\sum_{l=0}^{n} \frac{\lambda^{n-l}_{1}
\lambda^{l}_{2}}{(n-l)!l!}
(1+\lambda_{1})^{l-n}
(1+\lambda_{2})^{-l}\nonumber\\
\times
{\rm L}_{n-l}[-\frac{A_{1}}{\lambda_{1}(1+\lambda_{1})}]
{\rm L}_{l}[-\frac{A_{2}}{\lambda_{2}(1+\lambda_{2})}]\Big\rangle,
\label{12}\\
\begin{array}{lr}
=\sum_{l=0}^{n}\sum_{l_{1}=0}^{n-l}\sum_{l_{2}=0}^{l}
\sum_{l_{3}=0}^{n-l_{1}-l_{2}}\sum_{l_{4}=0}^{\infty}{n-l \choose l_{1}}
{l \choose l_{2}}{n-l_{1}-l_{2} \choose l_{3}} 4^{l_{1}+l_{2}+1}
\\
\\
\times \frac{(-1)^{n+l_{3}+l_{4}-l-l_{1}}
(n+l_{1}+l_{2}+l_{4}+1)! h^{l_{1}}_{1} h^{l_{2}}_{2}}
{l_{1}!l_{2}!l_{4}!(n+l_{1}+l_{2}+1)!}\exp [-\Gamma_{0}(t,l_{j})
+\frac{1}{2}\Gamma^{2}(t,l_{j})\sigma(t)]
, \label{12a}
\end{array}
\end{eqnarray}
where
\begin{eqnarray}
\begin{array}{lr}
\lambda_{1}=\frac{1}{2}[\exp(-2gt)- 1],\quad
\lambda_{2}=\frac{1}{2}[\exp(2gt)-1], \quad A_{1}=h_{1}\exp(-2gt),\\
\\
A_{2}=h_{2}\exp(2gt),\quad
h_{1}=\frac{1}{2}(|\alpha_{1}|^{2}+ |\alpha_{2}|^{2}+2|\alpha_{1}||\alpha_{2}|
\sin \psi),\\
\\
h_{2}=\frac{1}{2}(|\alpha_{1}|^{2}+ |\alpha_{2}|^{2}-2|\alpha_{1}|
|\alpha_{2}|\sin \psi),\\
\\
\Gamma_{1}(t)=-\frac{4(h_{1}-h_{2})tg_{1}\exp(-2g_{0}t)}
{[1+\exp(-2g_{0}t)]^{2}},\quad
\Gamma_{2}(t)=2h_{1}-\frac{2(h_{1}-h_{2})}{1+\exp(-2g_{0}t)},\\
\\
\Gamma(t,l_{j})=
\Gamma_{1}(t)+2g_{1}t(l_{1}+l_{2}+l_{3}+l_{4}+1),\\
\\
\Gamma_{0}(t,l_{j})= \Gamma_{2}(t)
+2g_{0}t(l_{1}+l_{2}+l_{3}+l_{4}+1)
\label{13}
\end{array}
\end{eqnarray}
and ${\rm L}_{n}(.)$ is the Laguerre polynomial of order $n$;
as before the angle brackets in (\ref{11}) and (\ref{12}) represent the
average over the stochastic pumping variable $\varepsilon$.
For simplicity
we have dropped the time $t$ from the coefficients $A_{j}$ and $\lambda_{j}$.
The quantities $A_{j}$ and $\lambda_{j}$ play the role of the mean numbers of
coherent photons and mean numbers of chaotic photons, respectively.
The transition from (\ref{11}) to
(\ref{11a}) as well as from (\ref{12}) to (\ref{12a}) has been done
by carrying out the integration over the stochastic variable.
It is reasonable  mentioning that the expressions
(\ref{11a}) and (\ref{12a}) are different from those in
\cite{in9,{in10}}, where the average over the initial complex field amplitudes
is taken there and consequently additional summations have appeared.
When performing this integration the following expressions
have been used in the treatment of the exponential factor
for the sum photon-number distribution:

\begin{eqnarray}
\begin{array}{lr}
\frac{A_{1}}{1+\lambda_{1}}
+\frac{A_{2}}{1+\lambda_{2}}=2h_{1}-2(h_{1}-h_{2})[1+\exp(-2gt)]^{-1}
\\
\\
=2h_{1}-2(h_{1}-h_{2})\sum_{l=0}^{\infty} (-1)^{l}\exp(-2lg_{0}t)
[1-2g_{1}tl\varepsilon +O(\varepsilon^{2})]
\\
\\
=2h_{1}-2(h_{1}-h_{2})\left[ \frac{1}{1+\exp(-2g_{0}t)}
+ \frac{2g_{1}tl\varepsilon
\exp(-2g_{0}t)}{[1+\exp(-2g_{0}t)]^{2}}\right]
,\label{A.4}
\end{array}
\end{eqnarray}
where in these relations $g$ is replaced by its value, and $O(\varepsilon^{2})$
includes terms of the power equal or higher than $\varepsilon^{2}$, which have been
neglected in the last step since $\varepsilon$ is small.
In spite of this the neglection is physically sensible and necessary for performing
the average over the parameter $\varepsilon$, it leads to  a problem that
the distribution (\ref{12a}) does not exist as an ordinary well behaved function.
Consequently additional treatments may be used to avoid this difficulty.
Therefore, if the propagation time through the crystal,
which is also the interaction time, is sufficiently short together with the
fact that the contribution of the correction part (the term involving
$g_{1}$) must be small compared with the main contribution, some modification
of the correction term can be performed.
For example we can approximate
$g^{2}_{1}t^{2}l_{4}^{2}$ by $g^{2}_{1}t^{2}(l_{1}+l_{2}+l_{3}+1)l_{4}$ in
 the exponential term.
 The choice of the factor $(l_{1}+l_{2}+l_{3}+1)$  is to compensate the
 large values of $l_{4}$ as much as possible since
$l_{j},j=1,2,3$
are the indices of the summations (cf. (\ref{12a})).
 It is important to point out for a short time interaction (as well as
finite values of $g_{1}$ and $\sigma_{0}$) and finite  values of $l_{4}$
that the contribution of
$\Gamma^{2}_{2}(t)$ is considerably small  compared with
that of $\Gamma_{0}(t)$.
The validity  of this approximation can be checked numerically.
Now the sum photon-number distribution (\ref{12a})
becomes well behaved function,  as it should be,
and   having the form
\begin{eqnarray}
\begin{array}{lr}
P(n,t)
=\sum_{l=0}^{n}\sum_{l_{1}=0}^{n-l}\sum_{l_{2}=0}^{l}\sum_{l_{3}=0}^{n-l_{1}-l_{2}}
{n-l \choose l_{1}}
{l \choose l_{2}}{n-l_{1}-l_{2} \choose l_{3}}
\frac{(-1)^{n+l_{3}-l-l_{1}} h^{l_{1}}_{1}
h^{l_{2}}_{2}4^{l_{1}+l_{2}+1}}{l_{1}!l_{2}!}
\\
\\
\times
\left\{1+
\exp \left[2\sigma(t)g^{2}_{1}t^{2}
\left(1
+\frac{\Gamma_{1}(t)}{g_{1}t}
+3(l_{1}+l_{2}+l_{3}+1)\right) -2tg_{0}\right]
\right\}^{-(n+l_{1}+l_{2}+2)}
\\
\\
\times\exp\left\{-2tg_{0}(l_{1}+l_{2}+l_{3}+1)-\Gamma_{2}(t)
+ \frac{1}{2}\sigma(t)[\Gamma_{1}(t)
+2tg_{1}(l_{1}+l_{2}+l_{3}+1)]^{2}\right\}.
 \label{13a}
\end{array}
\end{eqnarray}
One can easily check that when $g_{1}\rightarrow 0$, all equations for
fluctuating pumping  reduce to the nonfluctuating ones.

\begin{figure}
  {\includegraphics[width=8cm]{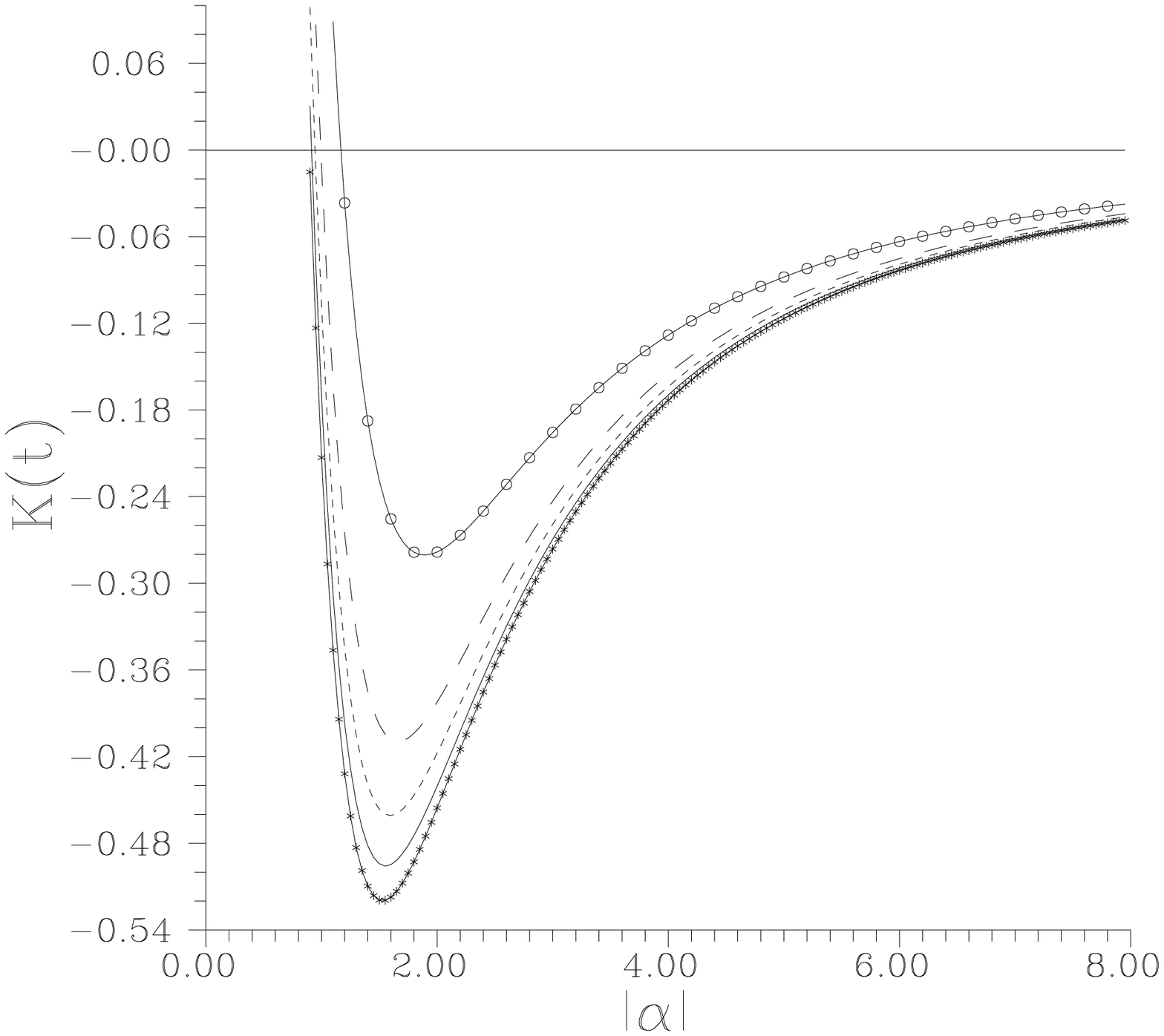}}
\caption{The normalized correlation of fluctuations between
modes\newline
 $[K(t)=\langle :\triangle W_{1}\triangle W_{2}:\rangle
/\langle W_{1}\rangle \langle W_{2}\rangle-1]$,
against $|\alpha |$
for $\psi=\pi/2, t=0.5, g_{0}=1, \mu=1, g_{1}=0.6$,
for the time-dependent fluctuating pumping
with $\sigma_{0}$=0.1 (solid curve), 0.5 (long-dashed curve) and
for the time-independent fluctuating pumping
with $\sigma_{0}$=0.1 (short-dashed curve), 0.5 (circle-centered curve).
The star-centered curve is corresponding to  that of the usual parametric amplifier
and the straight line is the bound of anticorrelation.}
\end{figure}
\par

In the following we investigate the behaviour of the system on the basis
of the results of the present section.

\section{Discussion of the results}
The discussion of the results of the model under consideration will be
performed through  two parts.
In the first part we give an investigation for both the anticorrelation
 between modes and the reduced factorial moments.
However, the second part is devoted to the analysis of the sum photon-number
distribution. Also in this part we develop new states and discuss some of its
properties.

\subsection{ Anticorrelations and reduced factorial moments}

We start our discussion in this part by analyzing the anticorrelations between modes
in the system, which can show some interesting features. For usual parametric
amplifier (i.e. $g=g_{0}$), one can easily check that the extreme values of
(\ref{2cc}) are established at $\psi=\pm\pi/2$, i.e. the maximum value
is at $\psi=-\pi/2$ and  minimum value at $\psi=\pi/2$. Careful examination shows
that it is  always positive when $\psi=-\pi/2$, i.e. the behaviour is always
classical. The maximal nonclassical effect  can appear  when $\psi=\pi/2$.
In other words, the anticorrelation  is observable as a result of
coupling of modes.
Further, considering that $\alpha_{1}=
\alpha_{1}=\alpha,\quad t>0$, and $\psi=\pi/2$,
one can easily obtain that   anticorrelation occurs if the
relation  (\ref{2cc}) is negative and this provides
\begin{equation}
|\alpha|>\frac{1}{2}\exp(g_{0}t)\sqrt{\sinh (2g_{0}t)}; \label{15}
\end{equation}
if the equality holds this expression means that coherent light can be
recovered. This relation gives the range of $|\alpha|$ over which the
anticorrelation  can be obtained.
The right hand side of (\ref{15}) is monotonically increasing function
in the course of time and then the nonclassical effect can
be realized for a certain time  interval  after  switching on the interaction.
It can be shown that the maximum value of anticorrelation
between modes is $-\frac{1}{4}\exp( 4g_{0}t)\sinh^{2}(2g_{0}t)$, which
occurs at  $|\alpha|=\frac{1}{2}\exp( 2g_{0}t)\sinh(4g_{0}t)$. Also the
maximum value of correlation establishes  at $|\alpha|=0$.
\begin{figure}
  {\includegraphics[width=8cm]{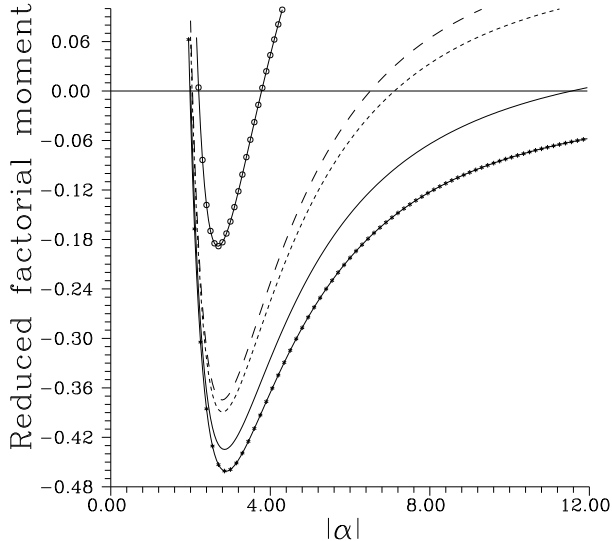}}
\caption{ The  reduced factorial moments $(\langle
W^{k}\rangle/\langle W\rangle^{k}-1)$ against $|\alpha |$ when
$k=5$ for $\psi=\pi/2, t=0.5, g_{0}=1, \mu=1, g_{1}=0.4$, for the
time-dependent fluctuating pumping with $\sigma_{0}$=0.1 (solid
curve), 0.3 (long-dashed curve) and for the time-independent
fluctuating pumping with $\sigma_{0}$=0.1 (short-dashed curve),
0.3 (circle-centered curve). The star-centered curve is
corresponding to  that of the usual parametric amplifier and the
straight line is the bound of antibunching.}
\end{figure}
Indeed, all these informations are clear in Figs. 1,2,
where the normalized forms of both  correlation between
modes and
reduced factorial moments have been plotted against $|\alpha |$,
respectively. From these figures one can see that the anticorrelation reflects
itself in the moments of reduced factorial moment. In general, the behaviour
of these quantities is smooth where, e.g. for reduced factorial moment, all curves  start from the bunching values going to the
antibunching ones  as the initial mean photon number increases until
they reach a maximum, then they start again to decrease the antibunching
values and eventually become stable for a large domain of $|\alpha|$.
The most effective antibunching  can  arise for the usual parametric
amplifier as expected (see star-centered curves).
When the pumping field fluctuations
are considered, the degradation of the nonclassical effect occurs
(see Figs. 1 and 2).
From these figures, the comparison between the solid and short-dashed
curves as well as between the long-dashed and circle-centered curves
shows that the nonclassical effects  for the standard
deviation in time-dependent case are more pronounced
than those for the time-independent case.
Further, one can see also that as the value of $\sigma_{0}$ increases, the
amount of antibunching decreases.

\begin{figure}
  {\includegraphics[width=8cm]{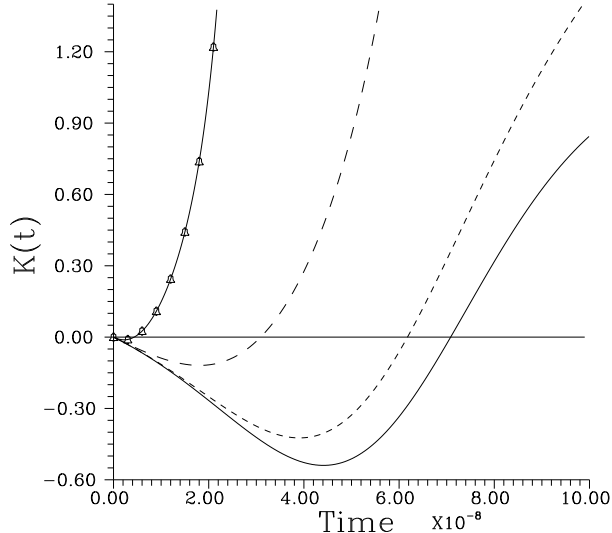}}
\caption{The normalized correlation of fluctuations between modes
for $\psi=\pi/2, |\alpha_{j}|^{2}=2, \mu =1, g_{0}=10^{7}s^{-1}$,
 for the time-dependent fluctuating pumping (solid curve) and for the
 time-independent fluctuating pumping.
Short-dashed curve is for $g_{1}=10^{6}s^{-1},\sigma_{0}=10$,
long-dashed curve for $g_{1}=10^{7}s^{-1},\sigma_{0}=1$,
bell-centered curve for $g_{1}=10^{7}s^{-1},\sigma_{0}=10$.
The solid curve is for the usual parametric amplifier as well as
for the time-dependent fluctuating pumping.}
\end{figure}

Now we would like to conclude this part by giving an example when the
interaction time is short which  is  relevant for practical situation
of propagating waves in nonlinear media.
We discuss this point only for the normalized correlation between modes
where similar arguments can be given for the reduced factorial moments.
This has been displayed in Fig. 3 for the shown values of the parameters.
We note when the standard  deviation is time-dependent, the behaviour of
the normalized correlation of fluctuations between modes
 is the same for all chosen values and
coincides with the behaviour of the usual parametric amplifier, which  gives
the most effective anticorrelation values (see solid curve in Fig. 3).
This seems to be related to the short interaction time.
Also we noted that the anticorrelation nature becomes less pronounced
as $g_{0}$ (or interaction time $t$) increases, where
the amplification of radiation can occur and this is connected with an
increase in chaotic noise \cite{in14}. Indeed, the regime of anticorrelation
or antibunching is related to the damping of the field \cite{in5,{in14}}.
In conclusion one can  control  the intensity of initial fields,
phases, and interaction time in such a way that we can get
maximum antibunching effect. This phenomenon can be observed by detecting
both the modes 1 and 2 simultaneously beyond the nonlinear medium.
It is reasonable to mention that  the sum of mode noises of $W=W_{1}+W_{2}$
can be less than modulus of negative values of $2\langle :\triangle W_{1}
\triangle W_{2}:\rangle$, i.e. antibunching can occur  if
$\langle :(\triangle W)^{2}:\rangle<0$ ($\langle :\triangle
W_{1}\triangle W_{2}:\rangle<0)$, which we have also verified. Such situation
cannot occur in classical physics and this leads to the nonexistence of the
Glauber-Sudarshan quasidistribution. Such quiet light may be applied in
optical signal processing and optical communications.

\subsection{ Sum photon-number distribution}

Firstly at $\alpha_{j}=0,\quad j=1,2$, for usual parametric amplifier,
 equation (\ref{12}) for the sum photon-number distribution reduces to
\begin{equation}
P(n,t)=\frac{[-\tanh (gt)]^{n}}{\cosh^{2} (gt)}\sum_{l=0}^{n}(-1)^{l}.
\label{16}
\end{equation}
This expression shows that only even terms survive and this is nonclassical
effect. Moreover, such photon-number distribution indicates that the
system collapses to produce the  state
\begin{eqnarray}
\begin{array}{lr}
|\Psi(t)\rangle =\frac{1}{\cosh (gt)}\sum_{l=0}^{\infty}
[\tanh (gt)]^{l}\exp(il\chi)|2l\rangle,\\
\\
 =\frac{1}{2\cosh (gt)}\sum_{l=0}^{\infty}
[\tanh (gt)
\exp(i\chi)
]^{\frac{l}{2}}[1+(-1)^{l}]|l\rangle,
\label{17}
\end{array}
\end{eqnarray}
where $\chi$ is a relevant phase.
In fact, state (\ref{17}) is quite different from both squeezed
vacuum states \cite{yun2} and single-mode of two-mode squeezed vacuum
states \cite{ba4}.   These two states are determined respectively by
\begin{equation}
|\Psi (t)\rangle_{1} =\frac{1}{\sqrt{\cosh (gt)}}\sum_{l=0}^{\infty}
[\frac{\tanh (gt)}{2}]^{l}\frac{\sqrt{(2l)!}}{l!}|2l\rangle,
\label{17a}
\end{equation}

\begin{equation}
\rho_{\rm I} (t) =\frac{1}{\cosh^{2} (gt)}\sum_{l=0}^{\infty}
[\tanh (gt)]^{2l}|l\rangle\langle l|,
\label{17b}
\end{equation}
where $\rho_{\rm I} (t)$ is the reduced density operator representing
the single-mode case of two-mode squeezed vacuum
states.

\begin{figure}
  {\includegraphics[width=8cm]{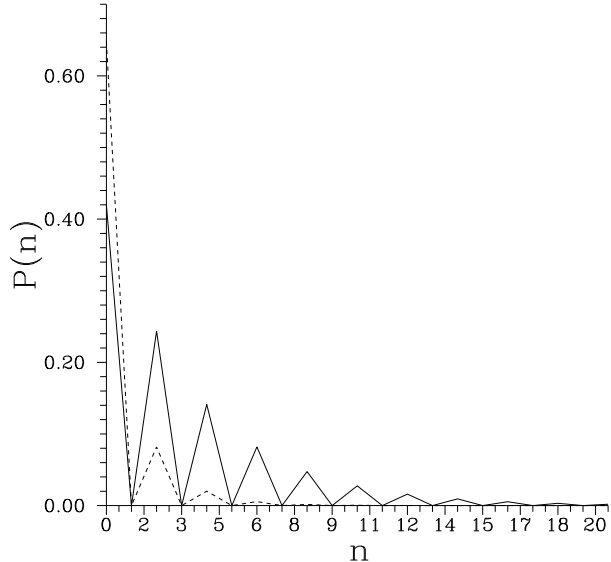}}
\caption{ The photon-number distribution of squeezed vacuum states
(dashed-curve) and of modified squeezed vacuum states (solid
curve) against the number of photons $n$ for $ t=0.5, g_{0}=2,
g_{1}=0$. }
\end{figure}
\par
The comparison of the three states (\ref{17})--(\ref{17b}) is instructive.
One can find that states (\ref{17}) are completely different from states
(\ref{17b}), where the latter is the state usually associated with chaotic
Bose-Einstein
statistics \cite{ba4}, i.e. it cannot exhibit nonclassical effects.
However, the former can exhibit nonclassical effects due to the superposition
principle. Actually, states (\ref{17}) can be written  in the form of
even thermal states by tracing over the phase through the relation
\begin{eqnarray}
\begin{array}{lr}
\hat{\rho}_{\rm eth}=\frac{1}{2\pi}\int^{2\pi}_{0}|\Psi (t)\rangle\langle
\Psi (t)|d\chi\\
\\
 =\frac{1}{\bar{n}+1}\sum_{l=0}^{\infty}\left(\frac{\bar{n}}{\bar{n}+1}
 \right)^{2l}|2l\rangle\langle 2l|, \label{18a}
 \end{array}
 \end{eqnarray}
where $\bar{n}=\sinh ^{2}(gt)$; in this case $\bar{n}$ represents the mean
number of thermal photons and $\hat{\rho}_{\rm eth}$ means the density
matrix of the even thermal states.
On the other hand,  states (\ref{17}) and (\ref{17a}) are different in
their structures but they can have similar nonclassical effects.
For instance, both states can have pairwise oscillation in the
photon-number distribution but with different amplitudes,
also they can display quadratures squeezing rather than sub-Poissonian statistics
(which  can be easily checked).
So that we shall call state (\ref{17}) as a modified squeezed vacuum state
(or even thermal state).
In Fig. 4 we have plotted the photon-number distribution for squeezed vacuum
and modified squeezed vacuum states for comparison.
From this figure  it is clear that the oscillations in the photon-number
distribution of state (\ref{17}) are more pronounced than those of (\ref{17a}).
These oscillations are purely quantum effect without classical analogue.
The origin of these pairwise oscillations in (\ref{17a}) is
 two-photon nature of the squeeze operator \cite{yun2} and
they have been interpreted by means of the interference in phase space \cite{sh}
or in the framework of the generalized superposition of coherent fields and
negative quantum noise \cite{bayer}. Nevertheless, for the state (\ref{17}) the
origin of the oscillatory behaviour in the photon-number distribution is the
quantum interference between the components of the state (see second line of
(\ref{17})) in phase space.
In fact, such a type of interference is quite different from the
interference  of the Schr\"{o}dinger cat states \cite{sch1}.
To make this point  clear we may analyze the behaviour of Wigner
($W$) function for modified squeezed vacuum states since this function is informative
and sensitive to the interference in phase space.
Applying standard technique, the $W$-function of (\ref{17}) is
\begin{equation}
W(z) =\frac{1}{\pi\cosh (gt)}\left\{
\exp\left[-gt-2|z|^{2}e^{-2gt}\right]+\exp\left[gt-2|z|^{2}e^{2gt}\right]
\right\},
\label{18}
\end{equation}
where $z=x+iy$.
\begin{figure}
 {\includegraphics[width=8cm]{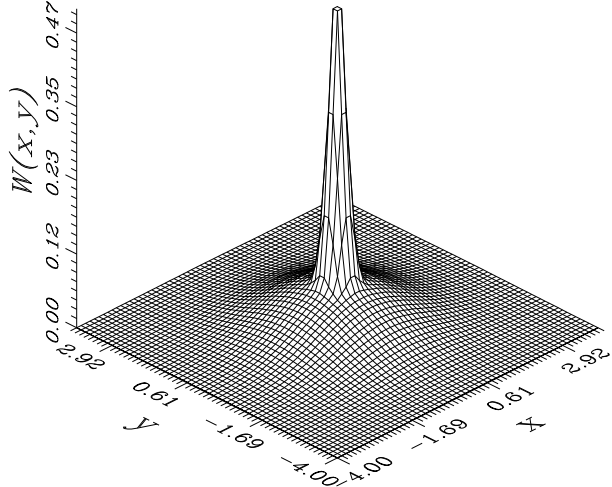}}
\caption{ The $W$-function of modified squeezed vacuum against
${\rm Re}z=x, {\rm Im}z=y$ for the same situation as Fig. 4. }
\end{figure}
\par

This relation shows that the $W$-function cannot exhibit negative values
and its contour in phase space is symmetric in contrast with that
of squeezed states showing  noise ellipse.
Further, it contains two terms, the first one is arising from the
statistical mixture and the second represents the interference between
the components of the state (\ref{17}).
Consequently the behaviour of
$W$-function here depends on the competition between these two terms
where both have a Gaussian shape with center at the origin, however,
the first term is always broader than  the second one.
This is clear in  Fig. 5 where $W$-function given by (\ref{18}) is plotted.
In fact, such  interference is responsible for the nonclassical effects of
the modified squeezed vacuum states.
Further, similar behaviour has also been seen
for even binomial states \cite{[9]}, which contains a finite
number  of Fock basis states.
Finally, it should be stressed here that states (\ref{17}) cannot be produced
by a unitary operator as the squeezed vacuum state.
This is so becuase there is no  Hermitian Hamiltonian  from which
the state (\ref{17}) can be generated via a canonical transformation.
 To be more specific, substituting
$|2l\rangle=\frac{\hat{a}^{\dagger 2}}{(2l)!}|2l\rangle$ into the right-hand
side of (\ref{17}), one can find that the summation cannot be
performed to obtain
the required unitary operator and this is in contrast with (\ref{17b}).
Further, the second line of (\ref{17}) shows that this state represents
a superposition of two like-phase-coherent states \cite{shap} and consequently
 it cannot be obtained
from any unitary operation \cite{knigh}.
\begin{figure}
{\includegraphics[width=8cm]{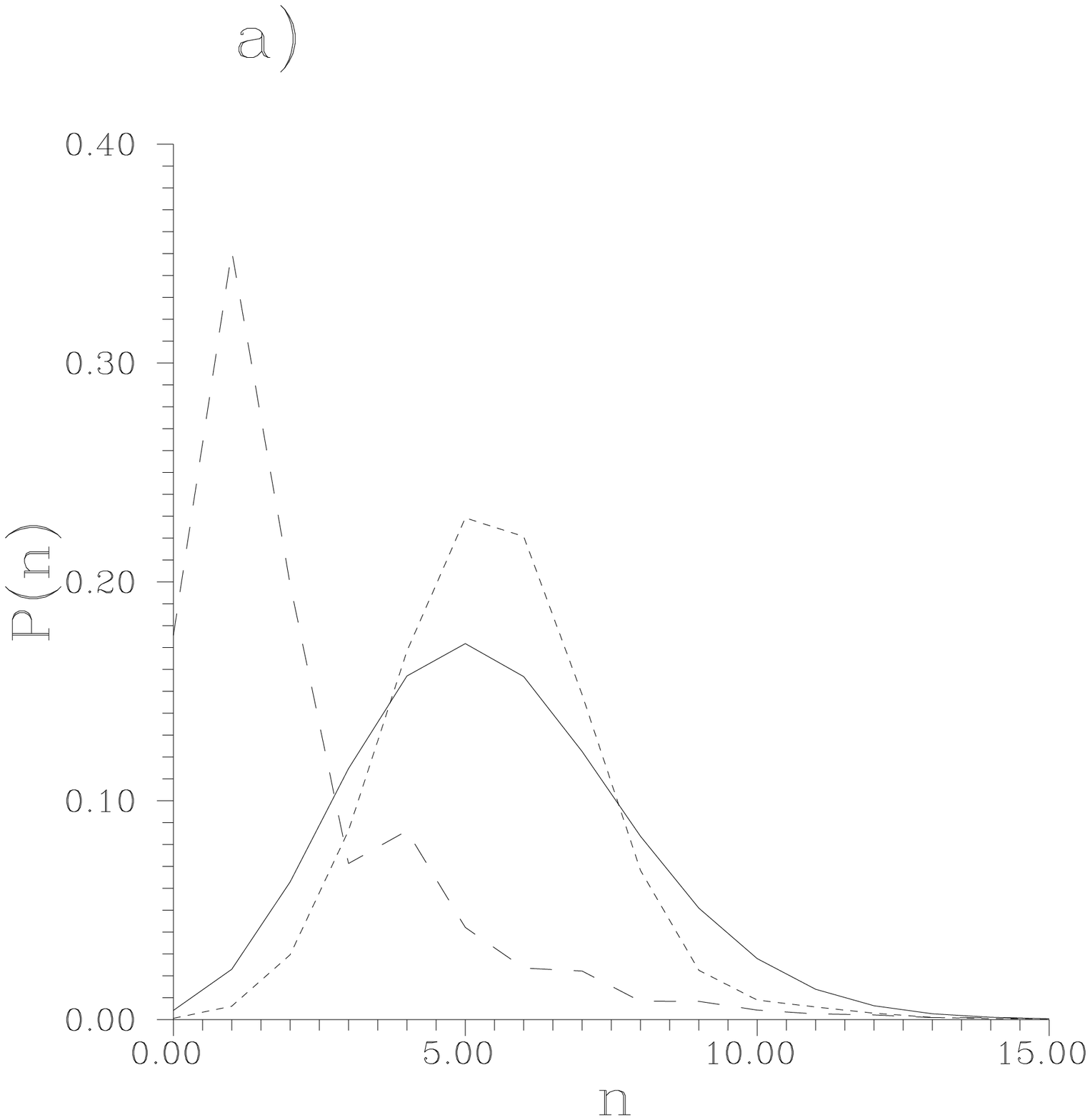}}
{\includegraphics[width=8cm]{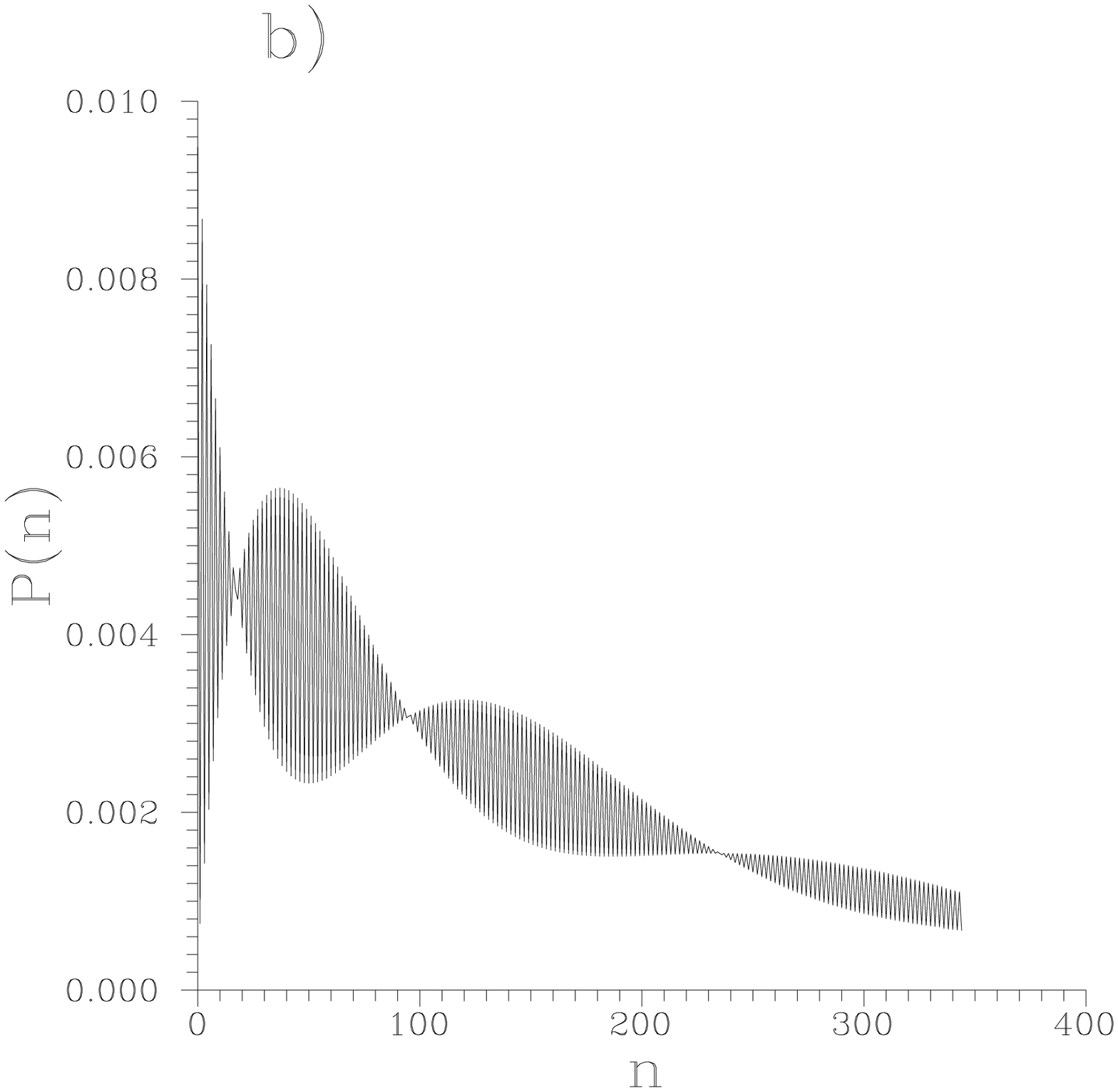}} \caption{ The sum
photon-number distribution against the number of photon $n$ for a)
usual parametric amplifier when $\psi=\pi/2, t=0.5,  g_{1}=0,\mu
=1, (g_{0}, |\alpha_{j}|)$= (1,2.5884) (short-dashed curve),
(1.5,0.87) (long-dashed curve), the solid curve is for
  the corresponding Poissonian distribution with the same
$\langle W(t)\rangle$;
b)
time-dependent fluctuating pumping when
$\psi=\pi/2, t=3\times 10^{-4},
 g_{1}=0.06,\sigma_{0}=0.02, (g_{0}, |\alpha_{j}|)=(10^{4},2)$. }
\end{figure}

 Now let us return  to the sum photon-number distribution
when $\alpha_{j}\neq 0,\quad j=1,2$.
Actually we noted that it is in a rough agreement with the
 normalized correlation of fluctuations between modes.
For instance, for the usual parametric amplifier and  when
$\psi=-\pi/2,\quad t>0$, $P(n,t)$ almost exhibits  single-peak structure
 broader than that of the corresponding coherent light with the
same mean photon numbers. However, when
$\psi=\pi/2$, $P(n,t)$ can exhibit  single-peak
or oscillatory behaviour according to the
relation between the incident mean photon numbers, coupling
constant and the interaction time
 (see Figs. 6a and b for shown values of the parameters).
In Fig. 6a  $|\alpha|=\frac{1}{2}\exp(2g_{0}t)\sinh(4g_{0}t)$, i.e.
when the maximum antibunching effects occur in normalized correlation
function between modes. One can observe that the distribution for this
case (short-dashed curve) is narrower than that of the corresponding
Poissonian distribution showing nonclassical effects.
Similar behaviours have been seen for the values of the parameters
satisfying inequality (\ref{15}).
On the other hand, when
$|\alpha|=\frac{1}{2}\exp(g_{0}t)\sqrt{\sinh (2g_{0}t)}-0.25$, i.e. the model
exhibits always bunching as it is clear from (\ref{15}). In this case
the photon-number distribution  can exhibit nonclassical oscillations
(long-dashed curve), however, it is broader than that for Poissonian light
(which is not included in the figure).
 The sum photon-number distributions of time-dependent and
time-independent fluctuating pump have a similar behaviour for short interaction time
(see Fig. 6b). From this figure there are evident the large scale macroscopic
oscillations in the behaviour of $P(n,t)$. To be more specific, $P(n,t)$ exhibits
collapses and revivals similar to those
familiar in the Jaynes-Cummings model (JCM) \cite{cum}; the former is in the photon
number domain rather than the time domain.
In fact, observations on the collapses and revivals of Rabi oscillations
were reported \cite{rempe}.  Further, collapses and revivals in the JCM are showing
the granular structure of the photon-number distribution, however, here they
indicate the establishment of strong correlation between the signal and idler
modes in the nonlinear crystal.  Indeed, similar phenomenon has been seen for
the photon-number distribution of single-mode
\cite{dutta} and two-mode \cite{mary} squeezed coherent states
with complex squeeze and displacement parameters. For the two-mode case
the phenomenon has been realized  for the diagonal, and joint
photon-number distribution having a fixed number of photons in one of
the modes, however, the case displayed here (Fig. 6b) for the sum
photon-number distribution is much richer than for the above distributions
 \cite{mary}.

\section{Conclusions}
As was shown earlier \cite{in5,{in9},{in10}} one can get antibunching
from the system described by (\ref{1}) at a short time and for  specific choice of
the parameters. Also such an effect can be obtained from parametric
amplification process with quantum pumping \cite{in4} regardless of the
initial phase condition. In this case the source of this effect is the
quantization of the pumping.
In this article we have analyzed the role of the initial mean photon
numbers on generated nonclassical light (antibunched light)
which has not been considered earlier \cite{in5,{in9},{in10},{in11}}
so that the overall quantum
efficiency for the effect can be enlarged. In other words, the degree of
antibunching can be controlled depending on the initial mean photon numbers.
Further, we have shown
that considering  time-dependent  pump fluctuations,
they decrease the rate of degradation of the nonclassical effects compared
with the time-independent case.
This is evident from the behaviour of the reduced factorial moments
and the normalized correlation of fluctuations between modes.
The behaviour of the sum photon-number distribution is in agreement with
the behaviour of reduced factorial moments
and normalized correlation of fluctuations between modes showing antibunching,
however, in some situations they can display nonclassical oscillations
 including their collapses and revivals.

We have shown also that we can generate a modified squeezed vacuum
states (or even thermal states) when the two-mode vacuum states evolve
in the present interaction,  thus detecting the sum photon-number
distribution for outgoing light using photodetectors.
These states have  somewhat similar behaviours as squeezed vacuum states,
however, the origin of the nonclassical effects for these states
is in the quantum interference between components of the state.
The photon-number distribution for such states exhibits more pronounced
oscillations than those of squeezed vacuum states for the same chosen
parameters.

\section*{Acknowledgement}

J. P. and F. A. A. E-O. aknowledge the partial support from the Projects
VS96028,  LN00A015 and Research Project CEZ: J 14/98 of Czech Ministry of Education
and from the Project 202/00/0142 of Czech Grant Agency.
One of us (M. S. A.) is grateful for the financial support
from the Project Math 1418/19 of the Research Centre, College of
Science, King Saud University.


\section*{References}
\begin{thebibliography}{200}

\bibitem{in1} W. H. Louisell and A. Yariv, Phys. Rev. 124 (1961)
1646.
\bibitem{kwia} P. G. Kwiat, W. A. Vareka, C. K. Hong, H. Nathel and
R. Y. Chiao, Phys. Rev. A  41 (1990) 2910; Z. Y. Ou, X. Y. Zou, L. J.
 Wang and L. Mandel, Phys. Rev. Lett. 65 (1990) 321.
\bibitem{zou1} X. Y. Zou, L. J. Wang and L. Mandel, Phys. Rev. Lett.  67
 (1991) 318.
\bibitem{in2} B. R. Mollow and R. J. Glauber, Phys. Rev. 160
(1967) 1076; 1097.

\bibitem{in3} L. Mi\v{s}ta, Czech J. Phys. B  19 (1969) 443.

\bibitem{in4} L. Mi\v{s}ta and J. Pe\v{r}ina, Acta Phys. Pol.
A 52(3) (1977) 425.
\bibitem{in5} L. Mi\v{s}ta and J. Pe\v{r}ina, Czech J. Phys. B 28
(1978) 392.
\bibitem {add} R. Vyas and S. Singh, Phys. Rev. A  40 (1989) 5147;
A. B. Dodson and R. Vyas, Phys. Rev. A 47 (1993) 3396.

\bibitem{in6}  W. S. Kryszewski and J. Chrostowski, J. Phys. A 10
(1977) 261.
\bibitem{in7} S. K. Srinivasan and S. Udayabaskaran, Opt. Act. 26
(1979) 1535.
\bibitem{in8}  W. J. Mielniczuk, Opt. Act. 26 (1979) 1115.
\bibitem{in9} V. Pe\v{r}inov\'a, Opt. Act.  28 (1981) 747.
\bibitem{in10} V. Pe\v{r}inov\'a and J. Pe\v{r}ina, Opt. Act.
28 (1981) 769.
\bibitem{in11}  J. Pe\v{r}ina: {\it Quantum Statistics of Linear
and Nonlinear Optical Phenomena}, 2nd edition (Kluwer, Dordrecht 1991).
\bibitem{in12} R. J. Glauber, Phys. Rev. 130 (1963) 2529.
\bibitem{in13} E. Schr\"{o}dinger, Naturwiss. 14 (1926) 664.
\bibitem{band} A. Bandilla and H.-H. Ritze, Opt. Commun. 34 (1980) 190.
\bibitem{ba1}  S. M. Barnett and P. L. Knight,  J. Opt. Soc. Am. B
2 (1985) 467.

\bibitem{ba4} S. M. Barnett  and P. L. Knight,  J. Mod. Opt.
34 (1987) 841.
\bibitem{in14} L. Mi\v{s}ta, V. Pe\v{r}inov\'a, J. Pe\v{r}ina
and Z. Braunerov\'a, Act. Phys. Pol. A 51 (1977) 739.
\bibitem{yun2} H. P. Yuen, Phys. Rev. A  13 (1976) 2226.
\bibitem{sh}  W. Schleich and J. A. Wheeler, Nature  326 (1987) 574;
 W. Schleich  and J. A. Wheeler, J. Opt. Soc. Am.
B  4 (1987) 1715; W. Schleich, D. F. Walls and J. A. Wheeler,
 Phys. Rev. A  38  (1988) 1177.

\bibitem{bayer} J. Pe\v{r}ina and J. Bajer,
 Phys. Rev. A 41 (1990) 516; J. Pe\v{r}ina, Z. Hradil and B. Jur\v{c}o,
 {\it Quantum Optics and Fundamentals of Physics} (Kluwer, Dordrecht 1994).
\bibitem{sch1}  E. Schr\"{o}dinger, Nature  23 (1935) 44.
\bibitem{[9]}  M. S. Abdalla, M. H. Mahran and A.-S. F. Obada, J.
Mod. Opt.  41 (1994) 1889;
  A.-S. F. Obada, M. H. Mahran, F. A. A. El-Orany and M. S. Abdalla,
 Int. J. Th. Phys.  35 (1996) 1393.
\bibitem{shap} J. H. Shapiro, S. R. Shepard and N. W. Wong,
 Phys. Rev. Lett. 62 (1991) 2377; J. H. Shapiro and S. R. Shepard,
  Phys. Rev. A 43 (1991) 3795.

\bibitem{knigh} K. W\'{o}dkiewicz, P. L. Knight, S. J. Buckle and S. M.
Barnett, Phys. Rev. A 35 (1987) 2567.
\bibitem{cum} E. T. Jaynes and F. Cummings, Proc. IEEE 51 (1963) 89;
N. H. Eberly, N. B. Norozhny and J. J. Sanchez-Mondragon, Phys. Rev. Lett.
44 (1980) 1323.
\bibitem{rempe} G. Rempe, H. Walther and N. Klein, Phys. Rev. Lett. 58 (1987)
353
\bibitem{dutta} B. Dutta, N. Mukunda, R. Simon and A. Subramaniam,
J. Opt. Soc. B 10 (1993) 253.
\bibitem{mary} M. Selvadoray, M. S. Kumar and R. Simon, Phys. Rev. A 49 (1994)
 4957.

\end{thebibliography}
\end{document}